\documentclass[twoside,twocolumn,prb,superscriptaddress]{revtex4-1}
\usepackage{hyperref}

\hypersetup{colorlinks=true, breaklinks=true, urlcolor=red, linkcolor=blue, bookmarksopen=true, citecolor=red}

\usepackage{comment}

\usepackage{graphicx}
\usepackage{dcolumn}
\usepackage{bm}
\usepackage{bbold}
\usepackage{amsmath}

\newcommand*{\vn}[1]{\bm{\mathrm {#1}}} 

\usepackage{graphicx}
\usepackage{placeins}

\usepackage{amssymb}
\usepackage[normalem]{ulem}
\usepackage{color}

\usepackage{perpage} 
\MakePerPage{footnote} 

\usepackage{multirow}
\usepackage{hhline}

\usepackage{float}
\usepackage{tabularx} 

\usepackage{array}
\newcolumntype{P}[1]{>{\centering\arraybackslash}p{#1}}

\usepackage{comment}

\usepackage[referable]{threeparttablex}
\usepackage{footnote}
\usepackage{hyperref}

\usepackage{threeparttable}

\newcommand{\fleur}{\texttt{FLEUR}}
\usepackage{makecell}
\usepackage{booktabs}

\renewcommand{\onlinecite}[1]{\nocite{#1}\citenum{#1}}

\newcommand{\comm}[1]{}

\begin{document}

\title{Ab initio analysis of magnetic properties of prototype B20 chiral magnet FeGe}

\author{S. Grytsiuk}
\email{s.grytsiuk@fz-juelich.de}
\affiliation{Peter Gr\"unberg Institut and Institute for Advanced Simulation, Forschungszentrum J\"ulich and JARA, 52425 J\"ulich, Germany}
\author{M. Hoffmann}
\affiliation{Peter Gr\"unberg Institut and Institute for Advanced Simulation, Forschungszentrum J\"ulich and JARA, 52425 J\"ulich, Germany}
\author{J.-P. Hanke}
\affiliation{Peter Gr\"unberg Institut and Institute for Advanced Simulation, Forschungszentrum J\"ulich and JARA, 52425 J\"ulich, Germany}
\author{P. Mavropoulos}
\affiliation{Department of Physics, National and Kapodistrian University of Athens, 15784 Zografou Campus, Athens, Greece}
\author{Y. Mokrousov}
\affiliation{Peter Gr\"unberg Institut and Institute for Advanced Simulation, Forschungszentrum J\"ulich and JARA, 52425 J\"ulich, Germany}
\author{G. Bihlmayer}
\affiliation{Peter Gr\"unberg Institut and Institute for Advanced Simulation, Forschungszentrum J\"ulich and JARA, 52425 J\"ulich, Germany}
\author{S. Bl\"ugel}
\affiliation{Peter Gr\"unberg Institut and Institute for Advanced Simulation, Forschungszentrum J\"ulich and JARA, 52425 J\"ulich, Germany}

\date{\today}

\begin{abstract}
FeGe in the B20 phase is an experimentally well-studied prototypical chiral magnet exhibiting helical spirals, skyrmion lattices and individual skyrmions with 
a robust length of 70~nm. While the helical spiral ground state can be verified by first-principles calculations based on density functional theory, this feature size could not be reproduced even approximately. 
To develop a coherent picture of the discrepancy between experiment and theory, we investigate in this work the magnetic properties of FeGe from first-principles using different electronic-structure methods. 
We study atomistic as well as micromagnetic parameters describing exchange and Dzyaloshinskii-Moriya interactions, and discuss their subtle dependence on computational, structural, and correlation parameters. In particular, we quantify how these magnetic properties are affected by changes of the lattice parameter, different atomic arrangements, exchange and correlation effects, finite Fermi-function broadening, and momentum-space sampling.
In addition, we use the obtained atomistic parameters to determine the corresponding Curie temperature, which agrees well with experiments. 
Our results indicate that the well-known and well-accepted relation between the micromagnetic parameters and the period of the helical structure, is not valid for FeGe. This calls for new experiments exploring the relation by measuring independently the spin stiffness, the spiralization and the period of the helical spin spiral.
\end{abstract}

\maketitle

\section{Introduction}

Magnetic noncentrosymmetric cubic crystals of B20-type, such as transition-metal germanides and silicides, are the class of materials for which the direct observation of chiral magnetic skyrmions has been reported first~\cite{Muhlbauer:2009, Yu:2010, Yu:2011, Munze:2010}.
Over the past decade, the study of these materials in bulk form~\cite{Lebech1989, Lee:2009, Wilhelm:2011, Dmitrienko12, Ritz:2013, Deutsch2014, Tanigaki2015, Shibata:2015, Kanazawa2016a, Shibata:2013, Grigoriev:2013, Altynbaev2016, Shibata2017, Siegfried2017}
or grown as films~\cite{Yu:2011, Karhu2012, Porter2014, Wilson14, Kanazawa2016, Turgut2018, Zheng2018} developed into an exciting research subject since they provide a perfect test ground for resolving fundamental properties of skyrmions, rendering them candidate materials for potential applications in skyrmion-based computation.
An important feature of these materials is the competition between the antisymmetric Dzyaloshinskii-Moriya~\cite{Dzyaloshinskii, Moriya:61} 
(DM) and the symmetric Heisenberg-type exchange interactions, resulting in a variety of striking magnetic phases with respect to temperature, magnetic field, material composition and geometry. At zero external magnetic field and below a critical temperature they are helimagnets. 
Most importantly, they exhibit typically a small pocket in the magnetic field versus temperature $(H, T)$ phase diagram, referred to as anomalous phase or the so-called A-phase~\cite{Lebech1989, Lebech1993, Grigoriev2009},
which has been identified with the skyrmion-lattice phase~\cite{Pfleiderer:2007, Muhlbauer:2009, Yu:2010, Yu:2011, Munze:2010, Wilhelm:2011}.
In addition to skyrmion lattices and single skyrmions, a more complex three-dimensional magnetic texture was observed for MnGe~\cite{Tanigaki2015}, and different types of topological excitations such as chiral bobbers may coexist with magnetic skyrmions in thin films of B20 compounds over a wide range of material parameters~\cite{Rybakov2015, Zheng2018, Adam2018}.

FeGe is the prototypical representative of the B20 compounds in which the A-phase~\cite{Lebech1989} and the bobber~\cite{Zheng2018} were first observed. The Curie temperature is close to room temperature and the helical period of about 70~nm is in a regime comfortable to resolve by experimental techniques. These are rather robust values confirmed by several experiments~\cite{Wappling1968, Lundgren68, Lebech1989, Siegfried2017, Turgut2018, Spencer2018}.
The most important micromagnetic parameters characterizing the magnetic order of B20 compounds are the spin stiffness $A$ and the spiralization 
$D$~\cite{Freimuth2014}, which define the helical period according to $\lambda=4\pi\, |A/D|$.
However, while the spin stiffness can be obtained from small-angle neutron scattering~\cite{Grigoriev2015} or magnetization behavior~\cite{Bloch, Dyson}, which yields values between $A=90\,$meV\AA$^2$ and $A=190\,$meV\AA$^2$ in FeGe~\cite{Siegfried2017, Turgut2018}, to our knowledge there are no direct independent experimental measurements of the spiralization constant $D$ in this compound.

In addition to these experimental studies, several theoretical efforts were undertaken to realistically model the magnetic properties of FeGe. Specifically, this includes micromagnetic and atomistic spin models based on Heisenberg-type exchange and DM interaction, where the underlying parameters are derived from first-principles calculations using density functional theory (DFT). 
However, in contrast to the robustness of the experimentally measured period of the magnetic modulations in FeGe, the theoretical micromagnetic parameters obtained by different techniques vary substantially.
For instance, theoretical predictions for the spiralization in FeGe yield $D=-4.5$~meV\AA{}~\cite{Gayles:2015} and $D=-6.5$~meV\AA{}~\cite{Spencer2018} based on the dispersion of spin spirals, whereas a relativistic multiple-scattering framework provides a value of $D=-9.0$~meV\AA{}~\cite{Mankovsky18}. Moreover, representing the spiralization by intrinsic spin currents leads to $D=-7.0$~meV\AA{}~\cite{Kikuchi:2016}, and theoretical studies focusing on the spin susceptibility report the two distinct values $D=-10.1$~meV\AA{}~\cite{Koretsune:2015} and $D=-1.0$~meV\AA{}~\cite{Koretsune:2018}. This large variation of the spiralization in FeGe is complemented by electronic-structure works that provide the values $A=700$~meV\AA{}$^2$~\cite{Koretsune:2018} and $A=855$~meV\AA{}$^2$~\cite{Kashin2018} for the spin stiffness, using the energy relation of non-collinear magnetic states or an approach based on Green's functions.

So far, the above mentioned DFT methods were not able to reproduce the experimentally observed period of the spin-spiral modulations in many B20 compounds.
Therefore, it is important to identify possible factors which might be very critical in computing the micromagnetic parameters by DFT.
In this work we focus on FeGe since spin fluctuations, which are difficult to catch with DFT, are much less relevant for this compound than for example in MnSi~\cite{Jeong:2004,Collyer:2008}. We explore by first-principles calculations based on different DFT methods how sensitive the micromagnetic parameters are to different factors, such as exchange-correlation potential, Hubbard-$U$ correction, broadening of the Fermi distribution, atomic position, and lattice parameter.
In addition, we gain microscopic insights by evaluating the atomistic parameters of Heisenberg and DM interactions, as well as the corresponding Curie temperature. We discuss the orientations of the DM interaction vectors with respect to the corresponding bonds and their contribution to the micromagnetic DM interaction following the symmetry of B20 materials.

The article is organized as follows. In Sec.~\ref{model} we introduce the theoretical spin model and provide the explicit relation between the atomistic and micromagnetic parameters of the exchange and DM interactions, focusing on the cubic B20 germanides. 
In Sec.~\ref{methodology} we briefly describe three different computational approaches and two electronic-structure frameworks which we employ in this work.
The computational details are summarized in Sec.~\ref{details}.
Section~\ref{results} presents our comprehensive analysis of the atomistic and micromagnetic interaction parameters in FeGe, where we discuss their dependence on structural details, computational parameters, and the choice of the electronic-structure method.
We conclude our work in Sec.~\ref{conclusions}.

\section{Magnetic Models}\label{model}

The magnetic interactions in B20 materials are typically modeled by a spin Hamiltonian
\begin{equation}\label{atom}
E= -\frac{1}{2}\sum_{i\neq j} J_{ij}{\vn S}_i\!\cdot\!{\vn S}_j 
-\frac{1}{2}\sum_{i\neq j}{\vn D}_{ij}\!\cdot\!\left[{\vn S}_i\!\times\!{\vn S}_j\right]\, ,
\end{equation}
where the microscopic parameters $J_{ij}$ and vectors ${\vn D}_{ij}$ describe the Heisenberg exchange and DM interactions, respectively, between classical spins ${\vn S}_i$ and ${\vn S}_j$ (treated as vectors with the length $|{\vn S}_i|=1$) of the magnetic atoms at different lattice sites $i$ and $j$. 
Here, we neglect the tiny magneto-crystalline anisotropy in the cubic B20 compounds~\cite{Lundgren1970}.
The B20 crystal belongs to the class of chiral crystal structures for which the orientation of the microscopic vectors $\vn D_{ij}$ can be arbitrary as they are not restricted with respect to the Fe-Fe bonds by the Moriya rules~\cite{Moriya:61}.

If the magnetic structure varies slowly across the crystal, i.e., $\vert\vn S_{j}-\vn S_i\vert/\vert \vn S_i\vert \ll \vert\vn R_{j}-\vn R_i\vert/a$, 
where $a$ is the lattice parameter and $\vert{\vn R}_i-{\vn R}_j\vert$ is the distance between atoms at sites $i$ and $j$ (${\vn R}_i \neq {\vn R}_j$), then a continuous magnetization vector field ${\vn m}({\vn r})$ (with $\vert {\vn m}({\vn r})\vert=1$) can be used to simplify the description of the magnetic properties. As a consequence, the entire effect of the exchange and DM interactions on magnetic structures can be summarized by introducing the spin-stiffness tensor $\mathcal{A}$ and the spiralization tensor $\mathcal{D}$~\cite{Schweflinghaus2016, Hoffmann2017} as micromagnetic parameters entering the generalized functional of the micromagnetic energy (defined per chemical unit cell)
\begin{equation}\label{eq:micro}
E[{\vn m}] = \frac{1}{V}
\int^V\!\text{d}{\vn r} \,
\bigg(
\nabla {\vn m}\,
\mathcal{A}\ 
\nabla {\vn m} + 
\mathcal{D}: \mathcal{L}({\vn m})
\bigg)\, ,
\end{equation}
where 
\begin{align}
\label{aa}
\mathcal{A} &= \dfrac{1}{4}\sum_{i\ne j} J_{ij} {\vn R}_{ij}\otimes{\vn R}_{ij}\, ,\\
\label{dd} 
\mathcal{D} &=\dfrac{1}{2} \sum_{i\ne j} {\vn D}_{ij}\otimes{\vn R}_{ij}\, ,
\end{align}
are contractions of the microscopic interaction parameters with the separation vector ${\vn R}_{ij} = {\vn R}_j - {\vn R}_i$.
Here, $\mathcal{D}: \mathcal{L}({\vn m}) =\sum_{\mu\nu}\mathcal D_{\mu\nu}\mathcal L_{\mu\nu}(\vn m)$ denotes the contraction 
with the chirality tensor $\mathcal{L}({\vn m}) = \nabla {\vn m} \times {\vn m}$, the components of which are the Lifshitz invariants of ${\vn m}$.
For more details, see Ref.~\onlinecite{Hoffmann2017}. 
The micromagnetic parameters, $\mathcal{A}$ and $\mathcal{D}$, defined by Eqs.~\eqref{aa} and~\eqref{dd}, are in units per chemical unit cell~\cite{note-units}, the index $i$ runs over all sites within the unit cell and $j$ runs over the whole lattice excluding pairs with $i=j$. In general, $\mathcal{A}$ and $\mathcal{D}$ are $3\times 3$ tensors. In practical calculations the sums in Eqs.~\eqref{aa} and~\eqref{dd} are truncated above a maximum interaction radius $R_\text{max}$.

The above tensors reduce to scalar matrices due to symmetry arguments as the cubic B20 materials are characterized by the point group $T$.
To make this point clear, we first group all symmetry-related pairs of atoms $i$ and $j$ into different shells with specific distances $R^s=\vert{\vn R}_{ij}\vert$. Then, summing up the outer products in Eqs.~\eqref{aa} and~\eqref{dd} over these symmetry-related pairs results in the expressions
\begin{equation}\label{At}
\begin{aligned}
\mathcal{A}
&=\dfrac{1}{4}
\sum_s^NJ^{s}
\begin{pmatrix} 
4{\vn R}^{s}\cdot{\vn R}^{s} & 0 & 0\\
0 & 4{\vn R}^{s}\cdot{\vn R}^{s} & 0\\
0 & 0 & 4{\vn R}^{s}\cdot{\vn R}^{s} 
 \end{pmatrix}\\
&= \mathcal{I}_3 
\sum_s^NJ^{s}\vert{\vn R}^s \vert^2 
=\mathcal{I}_3 \displaystyle \sum_s^N A^{s} = A\,\mathcal{I}_3\, ,
\end{aligned}
\end{equation}
and
\begin{equation}\label{Dt}
\begin{aligned}
\mathcal{D}
&=
\dfrac{1}{2}\displaystyle\sum_s^N
\begin{pmatrix} 
4{\vn R}^{s}\cdot{\vn D}^{s} & 0 & 0\\
0 & 4{\vn R}^{s}\cdot{\vn D}^{s} & 0\\
0 & 0 & 4{\vn R}^{s}\cdot{\vn D}^{s} 
 \end{pmatrix}\\
&= 2 \mathcal{I}_3 
\displaystyle\sum_s^N ({\vn D}^{s}\cdot{\vn R}^{s})
= \mathcal{I}_3 \displaystyle\sum_s^ND^{s}= D\,\mathcal{I}_3\, ,
\end{aligned}
\end{equation}
where ${\vn R}^s$, ${\vn D}^s$, and $J^s$ are representatives of the local bond properties ${\vn R}_{ij}$, ${\vn D}_{ij}$, and $J_{ij}$, respectively, within a given shell $s$. $N$ is the total number of considered shells and $\mathcal{I}_3$ is the identity matrix. The quantities $A^s$ and $D^s$ denote the contributions from shell $s$ to the spin stiffness and spiralization constants, respectively, which we define per chemical unit cell~\cite{note-units}. As follows from Eq.~\eqref{Dt}, each shell has the largest contribution to the DM interaction if ${\vn D}_{ij}^s \parallel {\vn R}_{ij}^s$, and it is zero if ${\vn D}_{ij}^s \perp\ {\vn R}_{ij}^s$.

\section{Computational Methodology}
\label{methodology}

In this work we employ three different computational approaches, briefly described in the following sub-sections, to extract the atomistic and micromagnetic interaction parameters from the electronic structure as determined by density functional theory. The approaches are distinct in the details, \textit{e.g.}, from which self-consistent state the parameters are extracted, which magnetic states are treated perturbatively, and at which stage of the calculation the spin-orbit coupling (SOC) is included. These computational frameworks are realized in two different electronic-structure methods that we present as well.

\subsection{Spin-spiral approach}
\label{SS:Spin-spiral_approach}

We assume a conical spin spiral characterized by the propagation vector ${\vn q}$ and the rotation axis $\hat{\vn e}_\text{rot}$. 
In general, the spin spiral of each magnetic atom type $\alpha=1,\ldots,4$ in the unit cell of FeGe can have an individual initial phase $\phi_\alpha$ and form a cone angle $\theta_\alpha$ with the rotation axis $\hat{\vn e}_\text{rot}$.
The orientation of any classical spin ${\vn S}_i={\vn S}_{n\alpha}$ at position ${\vn R}_{i}={\vn R}_{n\alpha}={\vn R}_n+\boldsymbol{\tau}_\alpha$ with $n$ labeling the unit cell at ${\vn R}_n$ and $\boldsymbol{\tau}_\alpha$ denoting the four magnetic sublattices within the unit cell, is described by
\begin{equation}\label{sp1}
{\vn S}_{n\alpha}({\vn q},\hat{\vn e}_\text{rot}) = 
\mathcal{R}(\hat{\vn e}_\text{rot})
\begin{pmatrix} 
\sin(\theta_\alpha)\cos({\vn q}\cdot {\vn R}_{n\alpha}+\phi_\alpha)\\
\sin(\theta_\alpha)\sin({\vn q}\cdot {\vn R}_{n\alpha}+\phi_\alpha)\\
\cos(\theta_\alpha)
 \end{pmatrix}\text{.}
\end{equation}
Here, $\vert {\vn S}_{n\alpha} \vert =1$ and 
$\mathcal{R}(\hat{\vn e}_\text{rot})$ is a unitary matrix mapping $\hat{\vn e}_3$ to the rotation axis $\hat{\vn e}_\text{rot}$ via the relation
$\mathcal{R}(\hat{\vn e}_\text{rot})\ \hat{\vn e}_3=\hat{\vn e}_\text{rot}$.

Experimental data~\cite{Lebech1989} and theoretical analysis~\cite{Bak80} suggest that the magnetic ground state of cubic FeGe is a flat helical spin spiral.
In this case $\theta_\alpha=90^\circ$, $\phi_\alpha=0^\circ$, and $\hat{\vn e}_\text{rot}\parallel{\vn q}$, which simplifies Eq.~\eqref{sp1} to
\begin{equation}\label{sp}
{\vn S}_{n\alpha}({\vn q}, {\vn R}_{n\alpha}) = 
\left[ 
\hat{\vn n}_1 \cos({\vn q}\cdot {\vn R}_{n\alpha}) +
\hat{\vn n}_2 \sin({\vn q}\cdot {\vn R}_{n\alpha}) 
\right] \, ,
\end{equation}
where $\hat{\vn n}_1$ and $\hat{\vn n}_2$ are mutually orthogonal unit vectors with $\hat{\vn{e}}_\text{rot}=\hat{\vn n}_1\times \hat{\vn n}_2$.
Considering spin spirals with slow rotation (corresponding to small wave vectors ${\vn q}$), \textit{i.e.}, if ${\vn S}_{n\alpha}({\vn q},{\vn R}_{n\alpha})$ can be approximated by the continuous vector field ${\vn m}({\vn q},{\vn r})$,
the total energy~\eqref{eq:micro} has the form
\begin{equation}\label{et1}
E({\vn q},\hat{\vn e}_\text{rot}) = 
{\vn q}^T\mathcal{A}{\vn q} - 
[\hat{\vn e}_{\text{rot}}\cdot \mathcal{D}]^T{\vn q} = E_\text{ex} + E_\text{DM}\text{ .}
\end{equation}
{Using the symmetry-dictated shapes of the tensors $\mathcal A$ and $\mathcal D$, Eqs.~\eqref{At} and~\eqref{Dt},}
we can simplify the energy dispersion for the helical spin spiral in B20 compounds:
\begin{align}
\label{en1}
E_{\text{ex}} &=
{\vn q}^T\mathcal{A}{\vn q}= A {\vn q}^T \mathcal{I}_3 {\vn q}
=A q^2\ ,\\
\label{en2}
E_{\text{DM}} &=
-[\hat{\vn e}_{\text{rot}}\cdot \mathcal{D}]^T{\vn q}
= -D [\hat{\vn e}_{\text{rot}}\cdot \mathcal{I}_3]^T{\vn q}=
-D \, q\text{ .}
\end{align}
where for helical spin spirals $\hat{\vn e}_{\text{rot}}\cdot{\vn q}=q$. 
Therefore the total energy of the magnetic interactions in B20 materials has the form 
\begin{equation}\label{energy}
E(q) = A q^2 - D q\text{ .}
\end{equation}
The wave number
$q_{\text{min}}=-{D}/{2A}$ that minimizes $E(q)$ defines the wavelength 
\begin{equation}
\lambda = \frac{2\pi}{\lvert q_{\text{min}}\rvert}=
4\pi\left|\frac{A}{D}\right|
\label{eq:lambda}
\end{equation}
of the spin spiral and its rotational sense as encoded in the sign of $D$.

According to Eq.~\eqref{energy}, the parameters $A$ and $D$ are related to derivatives of the total energy with respect to the wave number $\vn q$ in the long-wave length limit:
\begin{equation}
\label{en1a}
A =
\dfrac{1}{2}\left.
\frac{\text{d}^2E_{\text{ex}}({\vn q})}{\text{d} q^2}
 \right\vert_{q\rightarrow0}
\text{\quad and\quad }
D
=
-\left.
\frac{\text{d}E_\text{DM}({\vn q})}{\text{d} q}
\right\vert_{q\rightarrow0}\text{ .}
\end{equation}

\subsection{Infinitesimal rotation approach}
\label{SS:Inf-rot_approach}

An alternative route towards the parameters $A$ and $D$ is to determine first their microscopic origins via multiple-scattering theory as implemented in the Korringa-Kohn-Rostoker (KKR) Green's function method~\cite{kkr,kkr2, Papanikolaou}.
In this framework, the microscopic parameters of Heisenberg and DM interaction are obtained from the collinear state by applying infinitesimal rotations of the magnetic moments, which provides access to the atomistic parameters $J_{ij}$ and $\vn D_{ij}$~\cite{Liechtenstein:1987, Ebert:2009}. Based on this information, the micromagnetic analogs are found from Eqs.~\eqref{aa} and~\eqref{dd}. 
Since this formalism requires an integration over all occupied electronic states, varying the position of the Fermi level provides insights into the response of the magnetic interactions to doping and alloying~\cite{Zimmermann2018}.

\subsection{Berry phase approach}
\label{SS:Berry_phase_approach}

The recently developed Berry phase theory of DM interaction~\cite{Freimuth2014} constitutes a further conceptual and computational 
framework that allows us to evaluate directly the spiralization tensor $\mathcal D$ as linear response of the spin-orbit dependent free-energy density with respect to small chiral perturbations, based on the ferromagnetic state.
Bypassing non-collinear calculations, this approach facilitates the self-consistent treatment of the full spin-orbit interaction,
in contrast to the outlined method based on spin spirals.
The formalism correlates the DM interaction with the global properties of a mixed parameter space of the crystal momentum $\vn k$ and the magnetization direction $\hat{\vn m}$ according to the Berry phase expression~\cite{Freimuth2014} 
\begin{equation}\label{berry}
\mathcal D = \frac{1}{NV_\text{uc}} \sum_{\mu \nu} \hat{\vn e}_\mu \otimes \mathrm{Im} \sum_{{\vn k} n}^{\mathrm{occ}} \bigg[ \hat{\vn m} \times \bigg\langle \frac{\partial u_{{\vn k}n}}{\partial \hat{\vn m}} \bigg| h_{{\vn k}n} \bigg| \frac{\partial u_{{\vn k}n}}{\partial k_\nu}\bigg\rangle \bigg]\, .
\end{equation}
Here, $|u_{{\vn k}n}\rangle$ denotes an eigenstate of the lattice-periodic Hamiltonian $H_{{\vn k}}=e^{-i \vn k \cdot \vn r} H e^{i \vn k \cdot \vn r}$ with the band energy $\mathcal E_{{\vn k}n}$, $h_{{\vn k}n}= H_{\vn k} + \mathcal E_{{\vn k}n} -2 \mathcal E_\mathrm{F}$ where $\mathcal E_\mathrm{F}$ is the Fermi level,
$N$ is the number of $\vn k$-points, and the sum is restricted to all occupied states. 
By varying the number of occupied states, we can again access the response of the spiralization tensor due to doping or alloying.

\subsection{Electronic-structure methods}
\label{SS:Electronic-structure_methods}

The different approaches to extract first-principles interaction parameters from density functional theory are realized most efficiently in two different electronic-structure methods. 
The first one is the full-potential linearized augmented-plane-wave (FLAPW) method as implemented in the \fleur{} code~\cite{fleur}. The total energy of the spin spiral is calculated directly by applying the generalized Bloch theorem. 
We utilize this code to self-consistently compute the total energy in the absence of SOC for different values of the ${\vn q}$-vector. 
Thus, the variation of the electron density with the latter wave vector is included in the initial states from which we determine subsequently the energy of the DM interaction by treating SOC as a small perturbation~\cite{Heide2009}. Curvature and slope of the resulting two energy dispersions provide access to the micromagnetic parameters $A$ and $D$, respectively, see Eq. \eqref{en1a}.

The atomistic parameters of the magnetic interactions were computed using the full-potential relativistic Korringa-Kohn-Rostoker (KKR) Green's function method~\cite{kkr,kkr2, Papanikolaou} 
in which the all-electron charge density is obtained from the Green's function that is the solution of a Dyson-equation.
In contrast to the FLAPW method, the KKR method allows to compute the atomistic parameters $J_{ij}$ and ${\vn D}_{ij}$ with SOC included self-consistently, however, deviations of the electronic structure from the ferromagnetic state are not included.

In addition, to evaluate the spiralization tensor $\mathcal D$ according to its Berry phase theory, we compute the electronic structure of ferromagnetic states with various orientations $\hat{\vn m}$ using the \fleur{} code~\cite{fleur}. Based on this information, we generate systematically a single set of so-called higher-dimensional Wannier functions~\cite{Hanke2015}. This computational scheme facilitates an efficient but accurate advanced Wannier interpolation~\cite{Hanke2015,Hanke2017,Hanke2018} of the complex parameter space that underlies the calculation of Eq.~\eqref{berry}.

\section{Computational Details}\label{details}

Our calculations employ two different approximations to the {\itshape a priori} unknown exchange-correlation functional of DFT. While the generalized gradient approximation (GGA)~\cite{Perdew96} provides structural data in very good agreement with the experiment, which we thus use in all calculations, we also consider the local density approximation (LDA)~\cite{Perdew92} to reveal the role of exchange and correlation effects for the magnetic properties of FeGe. Specifically, this concerns the evaluation of the microscopic interaction parameters within the KKR method. Analogously, by using the LDA+$U$ methodology, we assess how correlations in the Fe-$3d$ and Ge-$4p$ orbitals affect the underlying magnetic properties. Using the structural GGA data, we use the values $1.5$, $2.5$, and $3.5$~eV ($1.5$~eV) for the Coloumb $U$ for Ge-$4p$ (Fe-$3d$) orbitals as well as $J=0.5$~eV.

The FLAPW calculations are converged with a plane wave cut-off of 4.2~a.u.$^{-1}$ and $24\times 24 \times 24$ $\vn k$-points in the full Brillouin zone (BZ). The muffin-tin radii were chosen as 2.2~a.u.\ for both Fe and Ge. Using GGA, we perform the structural optimizations within the \fleur{} code and use the resulting parameters among all computational approaches.
To accurately evaluate $A$ and $D$ from the dispersion of spin spirals, the $\vn q$-sampling of the energy curve has to match with the 
grid of points in the electron crystal-momentum BZ.
Since the ground state of FeGe is a long range helical spin spiral with $q_\text{m}=0.009$~\AA$^{-1}$, the explicit energy calculation requires 
$V_\text{BZ}/q_\text{m}^3 \approx 150^3$ ${\vn k}$-points in the full BZ of volume $V_\text{BZ}$,
resulting in an increased computational burden. Therefore, we compute the micromagnetic parameters from the corresponding dispersion curves obtained for larger $q$-values, and test the convergence.
All KKR calculations are performed using $48\times48\times48$ ${\vn k}$-points in the full BZ. 
The energies of the spin spiral and the micromagnetic parameters are computed using Eqs.~\eqref{atom}, \eqref{aa}, and \eqref{dd}, respectively, for which the summation is truncated above a maximal interaction radius of $R_\text{max}=5a$, where $a$ is the lattice parameter. 
We applied the infinitesimal rotation approach in LDA.

Aiming at the spiralization tensor within the Berry phase theory, we calculate self-consistently the FLAPW electronic structure of the ferromagnetic state with $\hat{\vn m}$ along the $z$ direction and all other parameters as stated before. Based on the converged charge density, we invoke the magnetic force theorem~\cite{Bruno}
to compute wave functions and band energies on a coarse $8$$\times$$8$$\times$$8$ $\vn k$-mesh for $8$ different magnetization directions. This information is used to generate systematically a single set of $114$ higher-dimensional Wannier functions~\cite{Hanke2015} out of $202$ Bloch states, with the frozen window extending to $5$\,eV above the Fermi level. In a final step, we employ an advanced Wannier interpolation~\cite{Hanke2015,Hanke2017,Hanke2018} to evaluate the Berry phase expression~\eqref{berry} by integrating over a dense mesh of $128$$\times$$128$$\times$$128$ $\vn k$-points.

\section{Results and Discussion}\label{results}

\subsection{Crystal structure and magnetic properties}

\begin{figure}[t!]\center
 \includegraphics[width=\linewidth]{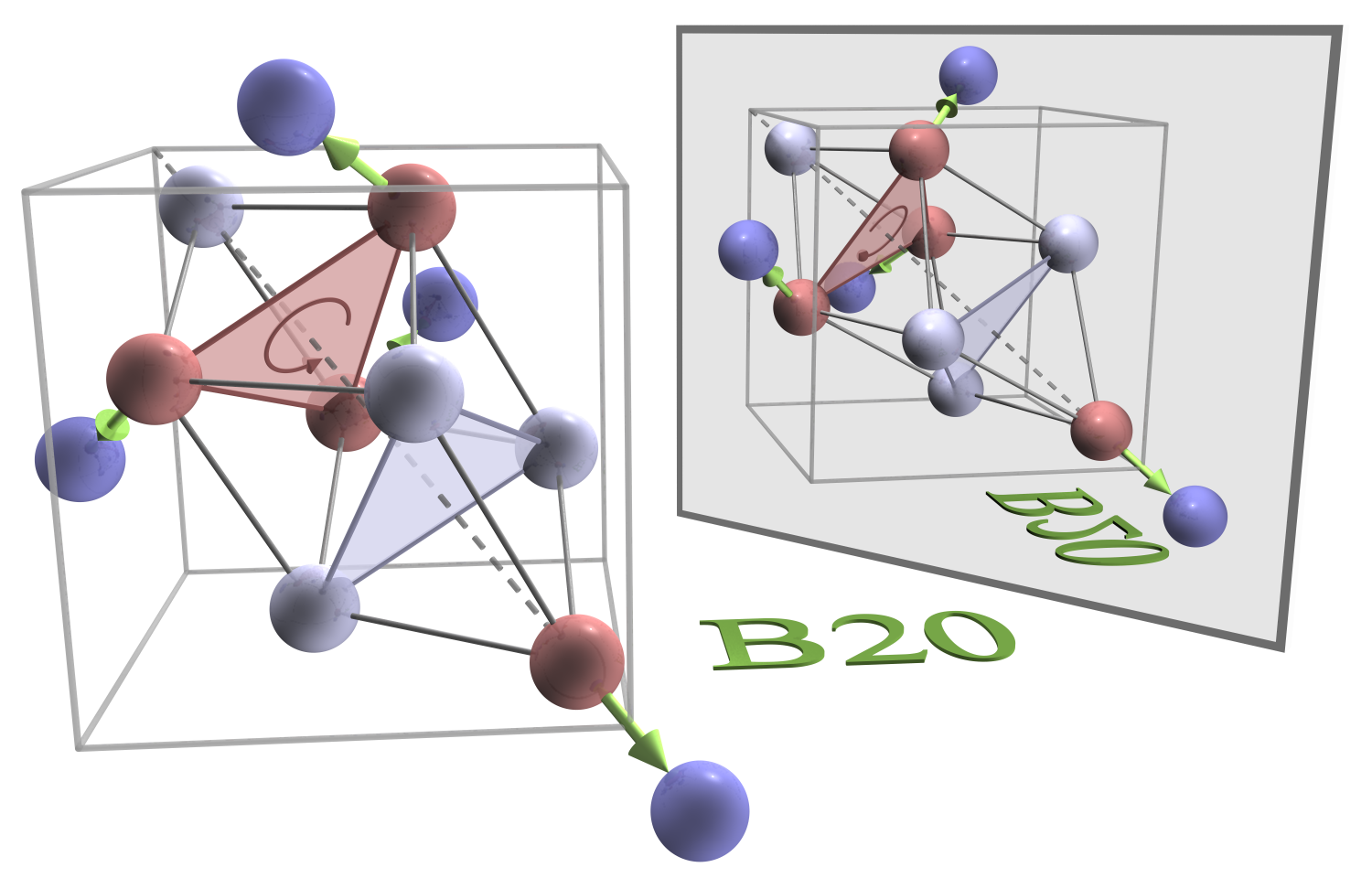}
 \caption{(Color online)
Visualization of the B20 structure. Shown is one of the enantiomers which can be transformed into the other by a mirror operation as illustrated. The red (light blue) spheres indicate the lattice positions of the magnetic (non-magnetic) ions located at the 4a Wyckoff positions $(u,u,u)$, $(0.5-u, 1-u, 0.5+u)$, $(1-u, 0.5+u, 0.5-u)$, and $(0.5+u, 0.5-u, 1-u)$. The corresponding quantities $u_\text{Fe}$ and $u_\text{Ge}$ for FeGe are given in Table~\ref{method}. To illustrate the structural chirality, the first-nearest non-magnetic neighbors of each of the four magnetic ions, positioned in the adjacent unit cells, are shown (dark blue spheres) additionally. They are located along the local three-fold rotation axes (green arrows).
} 
 \label{struc}
\end{figure}

The B20 structural type with the space group P$2_13$ of cubic FeGe does not contain symmetry operations of second kind and therefore corresponds to a chiral crystal structure. 
There are two sets of coordinates characteristic for two enantiomeric structures, one of which is shown in Fig.~\ref{struc}, that could be transformed one into another by inversion. 
In this work, we focus on the structure with right-handed crystalline chirality~\cite{Ishida1985}, which is defined by the 4a Wyckoff positions $(u,u,u)$, $(0.5-u, 1-u, 0.5+u)$, $(1-u, 0.5+u, 0.5-u)$, 
and $(0.5+u, 0.5-u, 1-u)$ for the magnetic and non-magnetic atoms.
The structural parameters obtained by DFT are in agreement with the experimental results in the temperature regime below 80~K, see Table~\ref{method}. The experimental saturation magnetic moment extrapolated to zero Kelvin at each Fe site is $\sim 1.0$~$\mu_\text{B}$~\cite{Wappling1968, Lundgren1970}.
A more recent experiment shows a magnetization of (360 $\pm$ 10)~kA/m at 5~K, which corresponds to a moment of (0.982 $\pm$ 0.007)~$\mu_\text{B}$~\cite{Spencer2018} per Fe atom. The calculated values obtained in DFT are slightly larger, 1.16~$\mu_\text{B}$ with GGA or 1.11~$\mu_\text{B}$ with LDA. 

The experimentally determined magnetic order of bulk FeGe in zero magnetic field is a long-period helical spin spiral with a period of about 70~nm, propagating along the crystallographic $[111]$ and $[100]$ directions at temperatures below 211 K and above 245 K,
respectively~\cite{Lebech1989, Ericsson81, Uchida08, Dussaux16}. The period of the helimagnetic order is very robust and remains unchanged also in thin films, although the propagation direction does not depend anymore on temperature and magnetic-field direction but it is normal to the film plane due to the change of the magnetic anisotropy~\cite{Spencer2018, Kanazawa2016}. 
Most of the theoretical studies on B20 compounds use a model of classical Heisenberg ferromagnetism with DM interaction~\cite{Bak80}, which we discuss in detail and compute using different models and DFT techniques, concentrating on different computational and structural factors that might have strong impact.

\newlength\q
\setlength\q{\dimexpr 0.065\textwidth -2\tabcolsep}
{\renewcommand{\arraystretch}{1.3}
\begin{table}[!b]
\centering
\caption{
Lattice parameter $a$ (in \AA), atomic positions $u_\text{Fe}$ and $u_\text{Ge}$, Fe-Ge and Fe-Fe distances (in \AA), and magnetic moment $S_\text{Fe}$ of the Fe atoms (in $\mu_\text{B}$). 
Experimental values as well as our DFT results obtained within GGA and LDA are provided.} 
\begin{threeparttable}
\begin{tabular}{l P{\q}P{\q}P{\q}P{\q}P{\q}P{\q}}\hline
 & a & $u_\text{Fe}$ & $u_\text{Ge}$ & $R_\text{Fe-Fe}$& $R_\text{Fe-Ge}$ & $S_\text{Fe}$\\ \hline\hline
Exp.\phantom{ww} & 4.69\tnotex{1}\phantom{i}$^\text{,}$\tnotex{2} & 0.135\tnotex{1} & 0.842\tnotex{1}
 & 2.881\tnotex{2} & 2.391\tnotex{2} & $\sim$1.0\tnotex{3}\\\hline
 GGA & 4.670 & 0.134 & 0.842 & 2.862 & 2.366 & 1.16 \\
 LDA & 4.558 & 0.136 & 0.841 & 2.795 & 2.321 & 1.11 \\ \Xhline{1.5\arrayrulewidth}
 \end{tabular}\label{method}
\footnotesize
\begin{tablenotes}[para]
 \item[1] \label{1} \phantom{i}Lebech \cite{Lebech1989}\phantom{we}
 \item[2] \label{2} \phantom{i}Wappling \cite{Wappling1968}\phantom{ee}
 \item[3] \label{4} Spencer~\cite{Spencer2018}
\end{tablenotes}
\end{threeparttable}
\end{table}}

\subsection{Micromagnetic interaction parameters}\label{micro}

\newlength\qa\newlength\qb\newlength\qc
\setlength\qa{\dimexpr 0.11\textwidth -1\tabcolsep}\setlength\qb{\dimexpr 0.14\textwidth -1\tabcolsep}\setlength\qc{\dimexpr 0.21\textwidth -1\tabcolsep}
{\renewcommand{\arraystretch}{0.5}
\begin{table*}[!t]
\caption{Magnetic moment $S$ of the Fe atoms, the spin stiffness $A$~\cite{note-units}, the spiralization $D$, the period $\lambda$ of the spin-spiral modulations, and $T_\text{C}$ for the B20 magnet FeGe, as obtained by experiment or by DFT calculations. 
The theoretical micromagnetic parameters are evaluated within different computational approaches. In addition to the values obtained in this work from spin spirals, Green's functions, and Berry phase theory, we list results from formalisms based on spin currents and spin susceptibility~\cite{Koretsune:2018}.
{For the spin-spiral approach, exchange and correlation effects are also treated within the {\itshape ad hoc} scaling GGA+$\alpha$ (with $\alpha=0.834$) or within GGA+$U$, where a Hubbard-$U$ of $1.5$~eV is used either only on the Ge $p$-orbitals or only on the Fe $d$-orbitals, and $J=0.5$~eV. }
}
\begin{threeparttable}
\begin{tabular}{l l P{\qa} P{\qa} P{\qb} P{\qa} P{\qc}} \Xhline{1.5\arrayrulewidth}
& & && & &\\[-0.7em]
Method & Approx. & $S~(\mu_\text{B})$ & $A$ (meV\AA$^2$) & $D$ (meV\AA) & $\lambda$ (\AA) & $T_\text{C}$~(K)\\& & && & &\\[-0.7em]\hline\hline
\multirow{9}{*}{Spin spirals \phantom{www}} & & && & &\\[-0.5em]
 & LDA & 1.11 & \comm{540}563 & $-$4.5, $-$4.7\tnotex{1} & \comm{1508}1557 & \\[-0.2em]
 & GGA & 1.16 & \comm{625}654, 650\tnotex{2} & $-$5.9, $-$5.5\tnotex{2}, $-$6.5\tnotex{3} & \comm{1331}1390, 1485\tnotex{2} & \\
 & GGA+U$_\text{Ge}$/U$_\text{Fe}$ & 1.16/1.56 & 657/649 & $-$5.7/$-$4.8 & 1447/1716 & \\
 & GGA+$\alpha$& 1.0 & 961 & $-$6.2 & 1955 & \\\Xhline{0.5\arrayrulewidth} & & && & &\\[-0.7em]
Inf. rotation \hspace{5pt} & LDA & 1.11, 1.12\tnotex{4} & 529, 855\tnotex{5} & $-$4.5, $-$9\tnotex{4} & 1477 & 310$\pm$10 (MC), 232\tnotex{5}\, (MF)\\
 & & && & &\\[-0.7em]
 & & && & &\\[-0.7em]
Berry phase & GGA/LDA & & & $-$6.5/$-$5.3 & & \\
Spin curr./susc. & GGA & & & $-$7\tnotex{6}\,/$-$1\tnotex{6} & & \\
\Xhline{0.5\arrayrulewidth} & & && & &\\[-0.5em]
Experiments& & 0.98 $\pm$ 0.01\tnotex{3} &89$\pm$8\tnotex{7}, 194\tnotex{8} & $-$1.6\tnotex{7} & 697\tnotex{7}, 700\tnotex{9} & 
\comm{280\tnotex{10}\,,} 
\comm{279\tnotex{11}\,,}
\comm{278\tnotex{12}\,,}
\ 280$\pm$2\!
\tnotex{3}\phantom{i}{$^,$}\!
\tnotex{10}\phantom{w}{$^,$}\!
\tnotex{11}\phantom{w}{$^,$}\!
\tnotex{12}\, \\ & & && & &\\[-0.5em]
\Xhline{1.5\arrayrulewidth}
\end{tabular}\label{method1}
 \footnotesize
        \begin{tablenotes}[para]
            \item[1]\label{1} Gayles~\cite{Gayles:2015}\phantom{ai}
            \item[2] \label{2}  Kikuchi \cite{Kikuchi:2016}\phantom{i}
            \item[3]\label{3} Spencer~\cite{Spencer2018}\phantom{i}
            \item[4] \label{4}  Mankovsky \cite{Mankovsky18}
            \item[5] \label{5}  Kashin~\cite{Kashin2018}\phantom{i}
            \item[6] \label{6}  Koretsune~\cite{Koretsune:2018}
            \item[]\label{13}   MC - Monte Carlo  
            \item[7] \label{7}  Turgut~\cite{Turgut2018}\phantom{ai}
            \item[8] \label{8}  Siegfried~\cite{Siegfried2017}\phantom{i}
            \item[9] \label{9}  Lebech~\cite{Lebech1989}\phantom{i}
            \item[10] \label{10}  Lundgren~\cite{Lundgren68}\phantom{i}
            \item[11] \label{11}  Xu~\cite{Xu2017}\phantom{ww}
            \item[12]\label{12} Wilhelm~\cite{Wilhelm11}\phantom{ww}
            \item[]\label{14} MF - Mean Field  
        \end{tablenotes}
\end{threeparttable}\end{table*}}

\begin{figure}[t!]\center
 \includegraphics[width=0.4\textwidth]{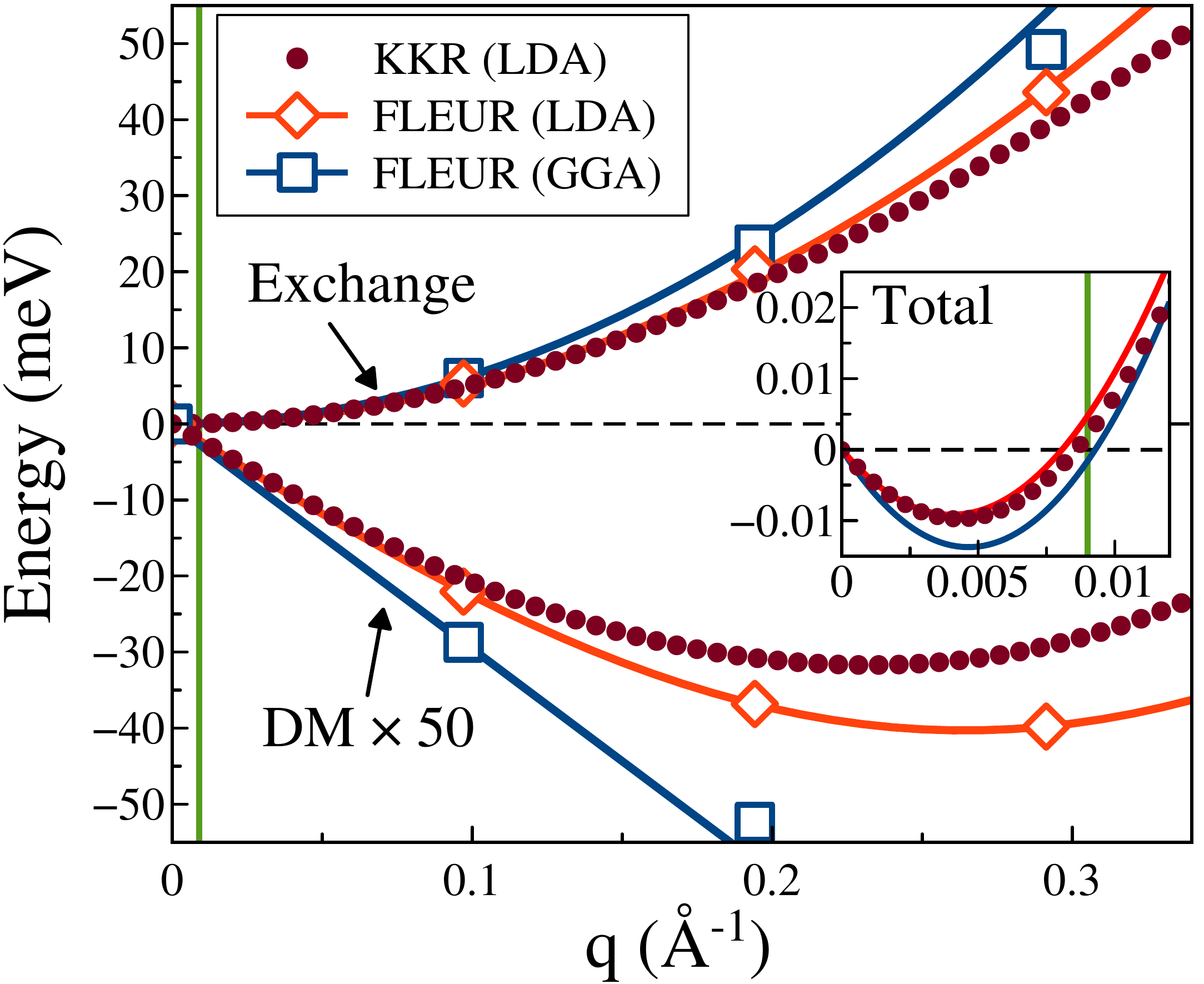}
 \caption{(Color online)
Energies per unit cell of the exchange and DM interaction, computed for flat spin spirals as function of $q=\vert{\vn q}\vert$ (with ${\vn q}$ pointing along the [111] direction). 
Please note the energy of DM interaction is multiplied by 50 for visualization purposes.
The inset shows the total energy around the minimum. Open markers show the energies computed by the \fleur{}
code (with LDA and GGA) and solid lines are fits to the corresponding energies 
$A q^2$, $D q$, and $A q^2 +D q$, respectively.
Filled circles are the energies obtained in accordance to Eqs.~\eqref{en1} and \eqref{en2}, where the pairwise parameters, $J_{ij}$ and ${\vn D}_{ij}$, were computed using the KKR method (with LDA).
The vertical green line represents the experimental pitch ($q_\text{exp}\approx 0.009$~\AA$^{-1}$).
} 
\label{fit}
\end{figure}

Figure~\ref{fit} displays the energy differences of the magnetic interactions (exchange and DM interaction) of a helical spin spiral with the wave vector ${\vn q}$ and the ferromagnetic state. Both dispersion curves were computed per unit cell either using the \fleur{} code (open markers) or from the microscopic KKR parameters $J_{ij}$ and ${\vn D}_{ij}$ (filled markers) entering Eq.~\eqref{atom}, as it is discussed in Sec.~\ref{SS:Electronic-structure_methods}. The full lines represent the analytical expressions for $E_\text{ex}(q) = A q^2$ and $E_\text{DM}(q) = D q$, where the parameters $A$ and $D$ are obtained through fits to the calculated energy points. The sum of both energies, $E_\text{ex}+E_\text{DM}$ is shown in the inset in the vicinity of the energy minimum for wave number $0\leq q\leq 0.01$~\AA$^{-1}$.

From Fig.~\ref{fit} it is clear that the energies of exchange and DM interactions 
agree rather well among spin-spiral and infinitesimal rotation approaches for small $q$-values and if the same exchange-correlation potential is used. Both LDA and GGA predict complex magnetic ground states characterized by similar wave vectors, {\itshape i.e.,} $q^\text{LDA}_\text{min}=0.0042$~\AA$^{-1}$ and $q^\text{GGA}_\text{min}=0.0047$~\AA$^{-1}$. 
The energy difference of about $0.01$~meV between the helical ground state and the FM state as obtained within LDA or GGA is tiny,
corresponding to a saturation magnetic field of $B=0.06$~T~\cite{Bfield}, which is twice as small as the experimental value obtained for FeGe~\cite{Bauer2016}.

The micromagnetic parameters obtained by the different computational approaches followed in this work, including spin-spiral dispersion, Lichtenstein formalism, and Berry phase theory, are summarized in Table.~\ref{method1}. 
The agreement between different electronic-structure methods improves if the same exchange-correlation functional is used.
In particular, the GGA results for the magnetic moment, the spin stiffness and spiralization amount to $S=1.16$~$\mu_\text{B}$, $A=654$~meV\AA$^{2}$, and $D=-5.9$~meV\AA, respectively, which is overall slightly larger than the values obtained within LDA leading to $S=1.11$~$\mu_\text{B}$, $A=563$~meV\AA$^{2}$, and $D=-4.5$~meV\AA, respectively.
The corresponding period $\lambda$ of the helical spiral amounts to 1560~\AA\ and 1390~\AA, respectively, which is consistently reproduced if we evaluate directly Eq.~\eqref{eq:lambda}. While the obtained $\lambda$ is similar to available theoretical data~\cite{Kikuchi:2016}, the value deviates by a factor of two from the experimentally measured pitch of $700$~\AA{} that corresponds to the wave vector $q^\text{exp}_\text{min}=0.009$~\AA$^{-1}$.

\begin{figure}[t!]\center
 \includegraphics[width=0.45\textwidth]{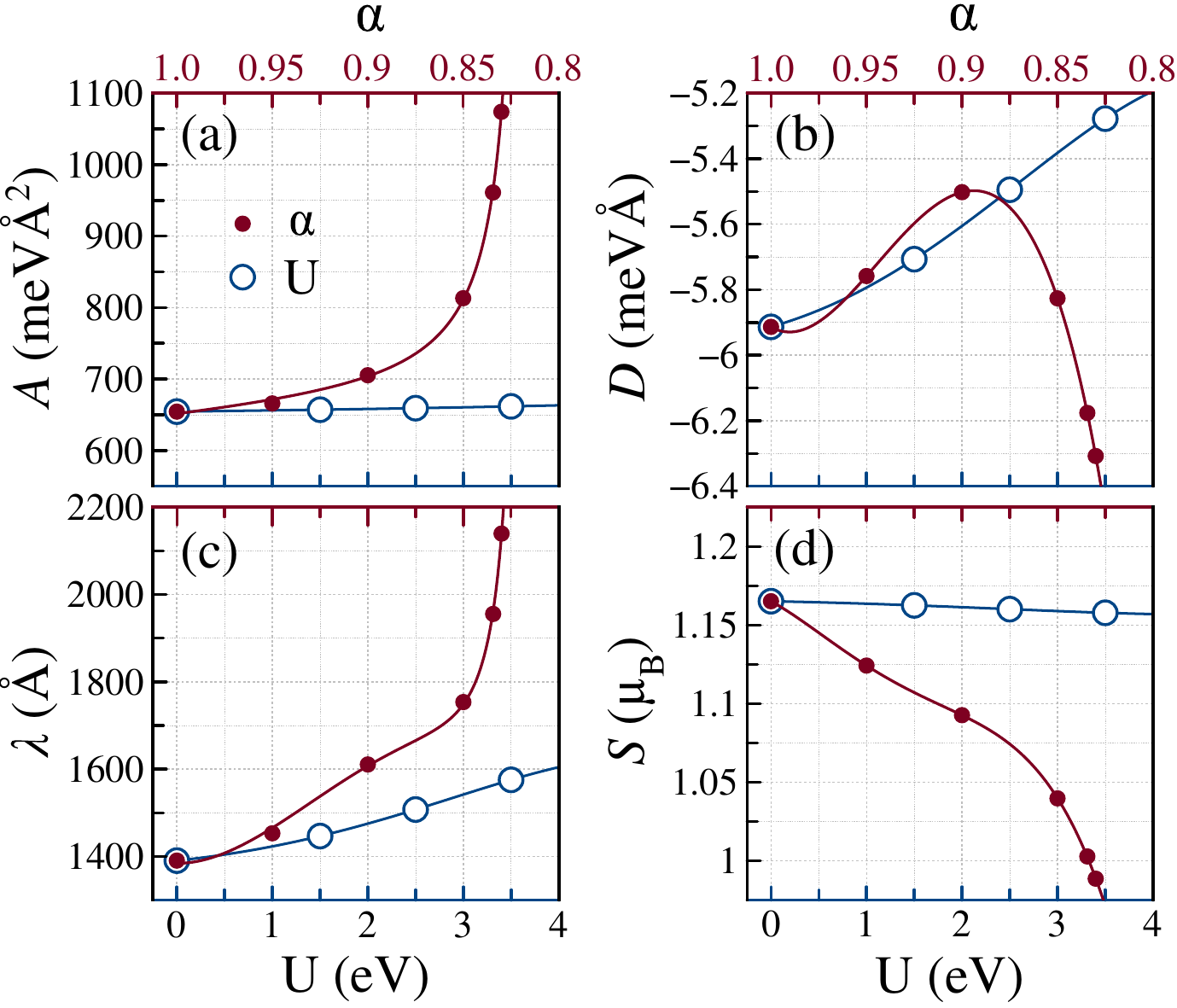}
 \caption{
 (a) Spin stiffness $A$, (b) spiralization $D$, 
 (c) period $\lambda$ of the spin spirals, and 
 (d) magnetic moment of the Fe atoms 
 as a function of the the Hubbard-$U$ parameter applied only to the $p$-orbitals of Ge 
 in combination with $J=0.5$~eV (open markers), and as a function of 
 the scaling factor $\alpha$ for the vector part of the exchange-correlation potential (filled markers).} 
 \label{ggaU}
 \end{figure}

This prominent discrepancy between experimental and theoretical wave vectors of the predicted magnetic ground state is unsatisfying. 
Therefore, we aim at analyzing the nature of this difference and trace it back to potential error sources, one of which could be the treatment of exchange and correlation effects in FeGe. To address this point, we follow an \textit{ad hoc} approach by scaling the vector part ${\vn B}_{\text{xc}}$ of the GGA exchange-correlation potential by a factor $\alpha$. 
For $\alpha=0.834$, this procedure reduces the magnetic moment to the experimentally determined value of about $1.0$~$\mu_\text{B}$, see Fig.~\ref{ggaU}(d). However, such a scaling enhances the spin stiffness ($A=961$~meV\AA$^{2}$) and the spiralization ($D=-6.2$~meV\AA{}), which manifests in an overall increase of the pitch to $\lambda=1955$~\AA{} that deviates even more from the experimental value.

To establish a clear picture of the role of correlations in FeGe, we apply a phenomenological Hubbard-$U$ correction either on the Fe-$d$ or on the Ge-$p$ states, on top of the GGA electronic structure. While introducing $U$ on the magnetic atoms has been reported to improve the agreement between theory and experiment on the magnetic moment in the B20 compound MnSi~\cite{Collyer:2008}, we demonstrate the opposite trend in FeGe. Using $U=1.5$~eV on the Fe-$d$ orbitals, the spin magnetic moment is strongly enhanced to $S=1.56~\mu_\text{B}$ as shown in Table~\ref{method1}, which is far from the measured value. In addition, due to such a correction the period of the spin spiral becomes longer by 23$\%$, reaching a value of 1716~\AA{}. On the other hand, using a Hubbard-$U$ to treat the correlations on Ge-$p$ orbitals might influence the hybridization of those states with Fe-$d$ orbitals and therefore could modify the strength of the magnetic interactions as well. As shown in Fig.~\ref{ggaU}, indeed, such an effect slightly reduces the moment, decreases spiralization, but enhances the spin stiffness such that the spin-spiral period becomes longer. 

From the above results we can conclude that exchange-correlation effects alone are not able to describe the discrepancy between the theoretically predicted and the experimentally measured length of the spin spiral in FeGe.
In addition, as shown in Table~\ref{method1}, there is a substantial variance in the micromagnetic parameters computed previously within different frameworks including infinitesimal rotation, spin current, and spin susceptibility.
Such discrepancy might be traced back to subtle computational details, the role of which we investigate in the following.

\subsection{Accuracy of the methods}

\begin{figure}[t!]\center
 \includegraphics[width=0.45\textwidth]{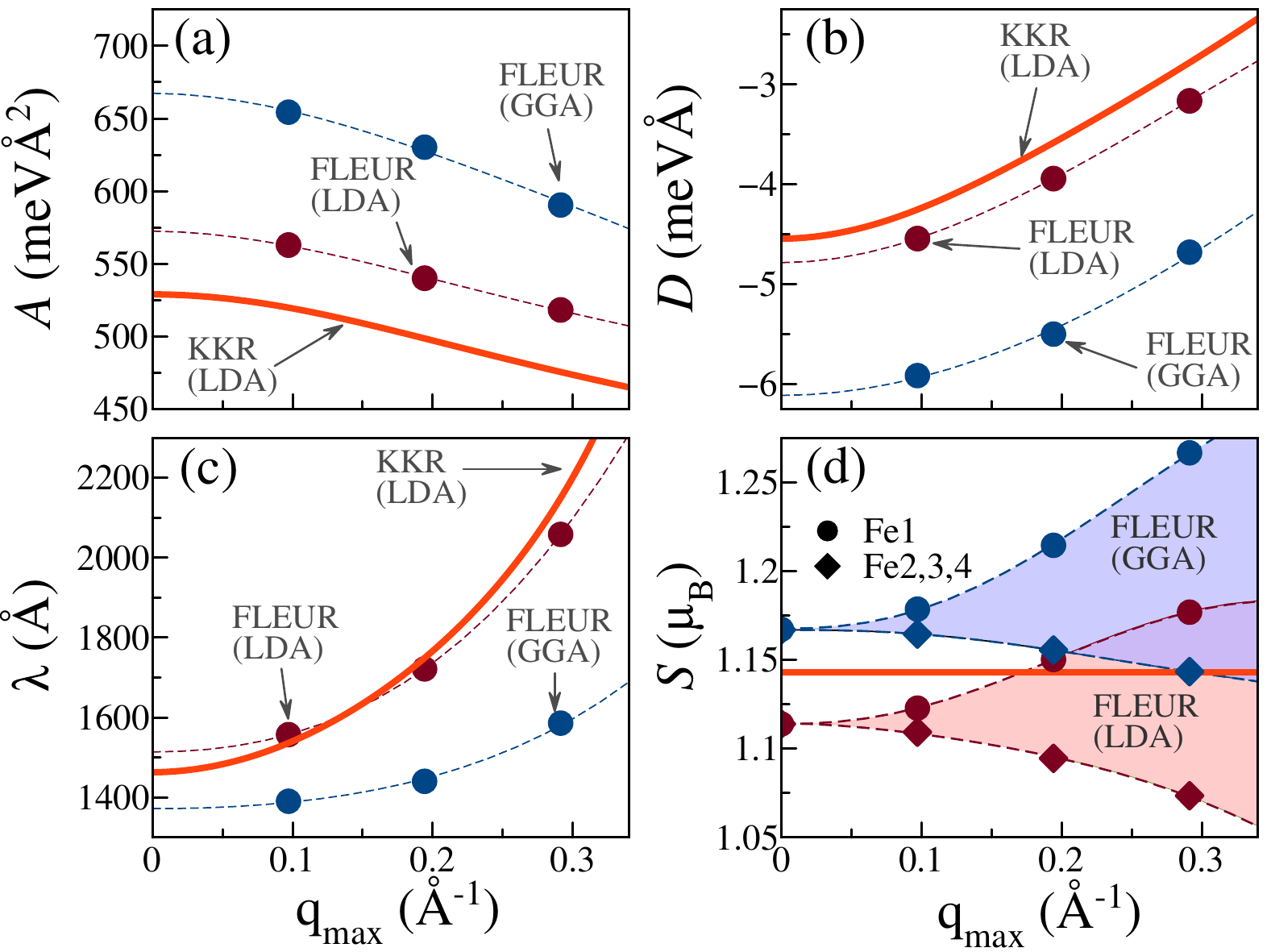}
 \caption{
 (Color online) Micromagnetic parameters of (a) the exchange $A$, (b) the DM interaction $D$ and (c) the corresponding period $\lambda=2\pi A/D$ of the resulting spin spiral, as a function of the fitting interval $[0,q_\text{max}]$.
 The micromagnetic parameters $A$ and $D$ were obtained as fit to the exchange and DM interaction energies of spin spirals shown in Fig.~\ref{fit}.
 (d) Magnetic moments of Fe atoms computed with the \fleur{} code for spin spirals with ${\vn q} \parallel{[111]}$ and with the KKR method for the FM state (${\vn q}=0$, red line). } 
\label{qmax}
\end{figure}%

In this section, we study the accuracy of our first-principles calculations of the micromagnetic parameters, focusing on the influence of fitting details and computational parameters such as sampling of the Brillouin zone (BZ) and broadening of the Fermi distribution.
First of all, since the micromagnetic parameters $A$ and $D$ within the spin-spiral approach are obtained 
by fitting the model expressions, Eqs.~\eqref{en1} and~\eqref{en2}, to the computed spin-spiral dispersions, 
we test the fit quality with respect to the size of the fitting interval $[0,q_\text{max}]$.
From Fig.~\ref{qmax}(a--c) it becomes clear that the absolute values of both $A$ and $D$ become larger if the fitting is performed for $q$-values closer to the $\Gamma$-point. Overall, the spin-spiral period hardly reduces below 
1460~\AA\ within LDA and 1370~\AA\ within GGA at $q=0$. 
Figure~\ref{qmax}(d) shows the magnetic moments of the different Fe atoms as obtained from self-consistent calculations without SOC, either of spin-spiral states in the \fleur{} code or of the ferromagnetic state in the KKR method. 
Note, different dependence of Fe magnetic moments on wave number $q$, as it is obtained within the \fleur{} code, is a result of broken symmetry by such magnetic structure. Minor differences of the magnetic moments at $q=0$ might be due to different integration range of the magnetization density within two methods. This might be the reason as well why the micromagnetic parameters $A$ and $D$ obtained within the two computational schemes are slightly different.

\begin{figure}[t!]\center
\includegraphics[width=0.45\textwidth]{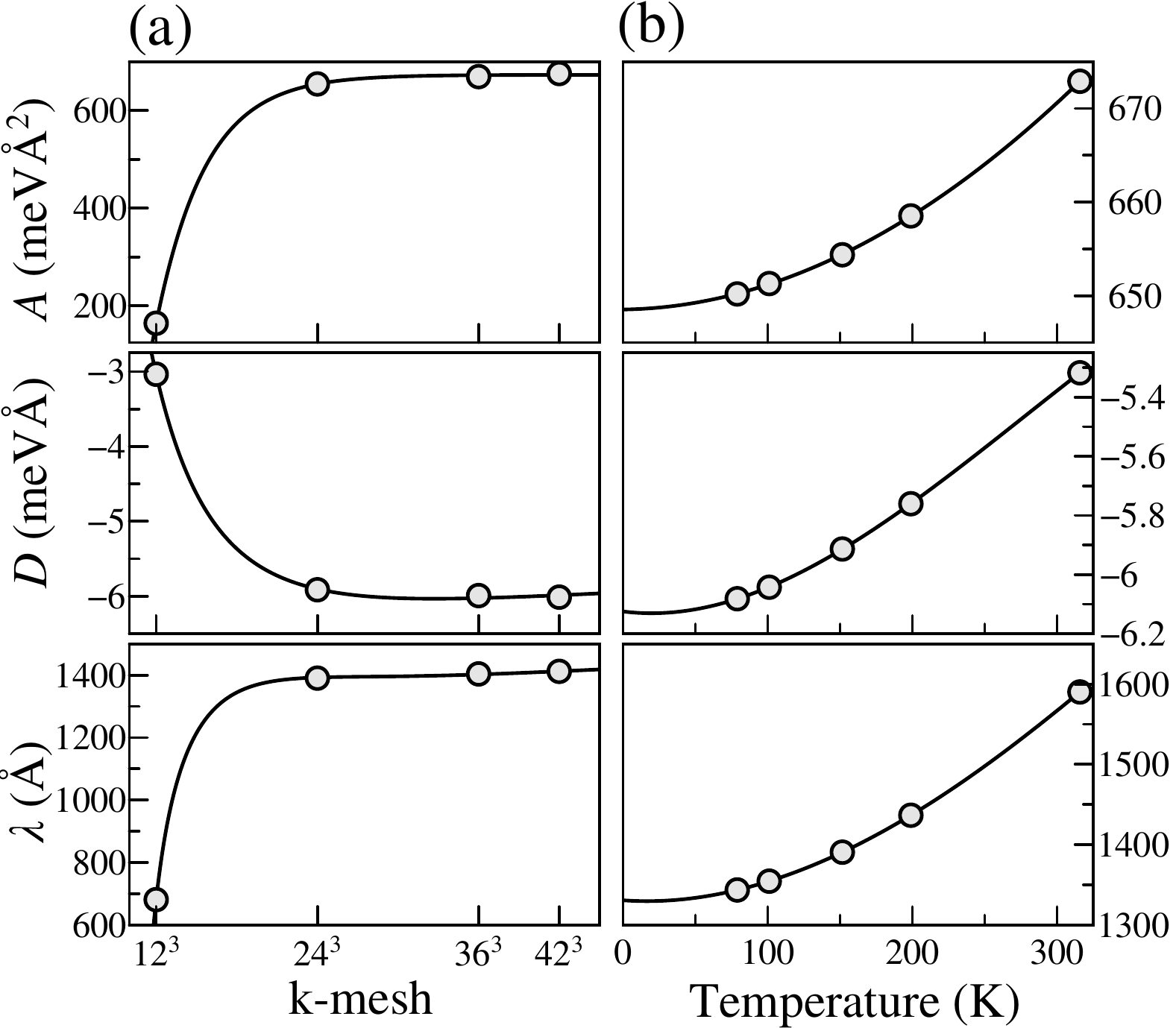}
\caption{
The spin stiffness $A$, the DM interaction $D$, and the spin-spiral period $\lambda$ as a function of (a) the number of $\vn{k}$-points in the full BZ and (b) the temperature of the Fermi smearing. Please note the different scales of the left and right axes.} 
\label{temp}
\end{figure}

Next we assess how the sampling of the BZ as well as the broadening of the Fermi distribution affect the convergence of the micromagnetic parameters. According to Fig.~\ref{temp}(a), the parameters $A$, $D$, and $\lambda$ are essentially converged to a robust value if we use more than $24\times24\times24$ $\vn k$-points. However, below this critical density for sampling momentum space, we recognize drastic changes in the micromagnetic parameters. For instance, if we use ${12\times12\times12}$ $\vn k$-points in the full BZ, spin stiffness and spiralization are four and two times smaller, respectively, resulting in a two times shorter period of the magnetic modulations $\lambda$.

For metallic systems the redistribution of the electronic states around the Fermi energy can play a key role for the magnetic properties~\cite{Gayles:2015, Koretsune:2018}. Therefore, it is important to investigate the effect of the Fermi broadening as mediated by the temperature $T$ on micromagnetic parameters. As shown in Fig.~\ref{temp}(b), lowering the temperature below 100~K does not change the micromagnetic parameters significantly.
Thus, it becomes obvious that the
computed wavelength of the helical state is closest to experiment if the Fermi broadening is small. A larger value of the Fermi broadening results in larger spin stiffness and smaller spiralization, manifesting therefore in a larger period of the spin spiral.

\subsection{Magneto-structural dependence}

\begin{figure}[!t]\center
 \includegraphics[width=0.45\textwidth]{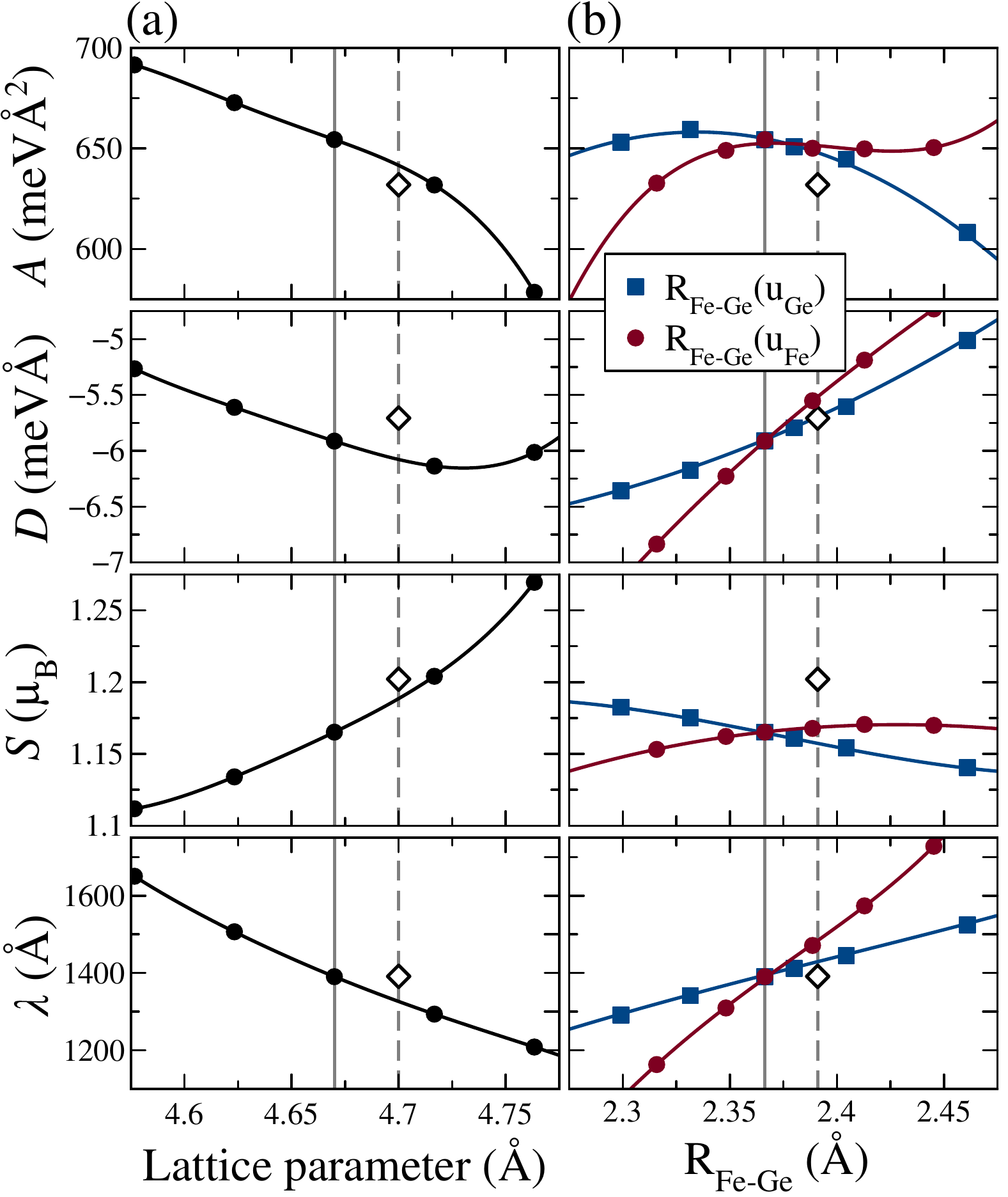}
 \caption{
 Spin stiffness, DM interaction, magnetic moment of Fe atoms, and 
 period of the spin spirals as function of (a) the lattice parameter $a$ and (b) 
 the distance between the first-nearest Fe and Ge atoms $R_\text{Fe-Ge}(u_\text{Fe},u_\text{Ge}) = \sqrt{3}(1 + u_\text{Fe} - u_\text{Ge})$.
 Filled markers in (a) stand for the structures with relaxed atomic positions for different lattice parameters while the open marker corresponds to a simulation with the experimentally obtained values.
 In (b) the distance between neighboring Fe and Ge atoms is shown either as function 
 of the Ge atomic position $u_\text{Ge}$ for $u_\text{Fe}=0.134$, or as a function of the Fe atomic position $u_\text{Fe}$ for $u_\text{Ge}=0.842$.
 The solid (dashed) vertical line stands for the relaxed (experimental) lattice parameter and atomic positions, see Table~\ref{struc}.
 } 
\label{lattice1}
\end{figure}

In this section, we study the dependence of the spin magnetic moment, the magnetic interaction parameters, and ultimately the wavelength of the helical state on the structural details such as lattice parameter $a$ and atomic positions ($u_\text{Ge}$ and $u_\text{Fe}$), considering two scenarios.
In the first case, the atomic positions were optimized for various lattice parameters while in the second case, $u_\text{Ge}$ and $u_\text{Fe}$ were tuned for a given lattice parameter.
The interaction parameters $A$ and $D$ that we present here were obtained from spin-spiral calculations based on the GGA ground state, the structural properties of which are in very good agreement with experiment.
For example, as summarized in Table~\ref{method}, the optimized lattice parameter of $a=4.67$~\AA{} is only $0.6$\% smaller than the experimental value, and also the relaxed atomic positions of Fe and Ge atoms hardly deviate from the measured values.

The micromagnetic parameters of FeGe as a function of the lattice parameter are illustrated in Fig.~\ref{lattice1}(a).
First of all, we note that upon increasing the lattice parameter, the distances between the first-nearest Fe-Ge
and Fe-Fe neighbors grow linearly, manifesting in a larger magnetic moment of the Fe atoms.
In addition, while increasing $a$ leads to a stronger DM interaction, the spin stiffness is reduced, as a consequence of which the period $\lambda$ increases.

Keeping the lattice parameter fixed to the equilibrium value $a=4.67$~\AA, next, we study how structural changes in terms of modified Fe and Ge positions affect the interaction parameters $A$ and $D$. During this analysis, we make use of the distance between neighboring Fe and Ge atoms, which amounts to $R_\text{Fe-Ge}=\sqrt{3}a_0(1+u_\text{Fe}-u_\text{Ge})$~\cite{comment1}. 
Remarkably, our results shown in Fig.~\ref{lattice1}(b) demonstrate that the micromagnetic
quantities $A$ and $D$ depend differently on $u_\text{Ge}$ and $u_\text{Fe}$ although $R_\text{Fe-Ge}$ is the same.
The spin stiffness $A$ is less sensitive to the positions of Ge and Fe atoms, whereas $D$ changes prominently with the Fe positions.
For example, reducing the Fe-Ge distance by $1\%$ by moving the Fe (Ge) atoms changes the spin stiffness only by $+1\%$ ($-0.5\%$), while the DM interaction is enhanced by $6.5\%$ ($2.8\%$).
For the spin stiffness, this behavior directly correlates with the trends for the spin magnetic moment shown in Fig.~\ref{lattice1}(b). 
Moreover, the increasing magnitude of the DM interaction can be attributed to 
the change in the hybridization between the orbitals of two atoms (Fe and Ge), as well as to an enhancement of the gradient of the electrostatic potential as Fe and Ge approach each other. 
Owing to these characteristics of the interaction parameters, the period $\lambda$ of the magnetic modulations reduces with decreasing distance $R_\text{Fe-Ge}$.

\subsection{Atomistic interaction parameters}
\label{subsection:AIP}

In the following, we elucidate the microscopic nature of the magnetic interactions in FeGe by discussing the atomistic exchange parameters obtained within the KKR formalism. Specifically, we analyze their contributions to the micromagnetic quantities describing Heisenberg and DM interactions. 

\begin{figure}[t!]\center
 \includegraphics[width=0.45\textwidth]{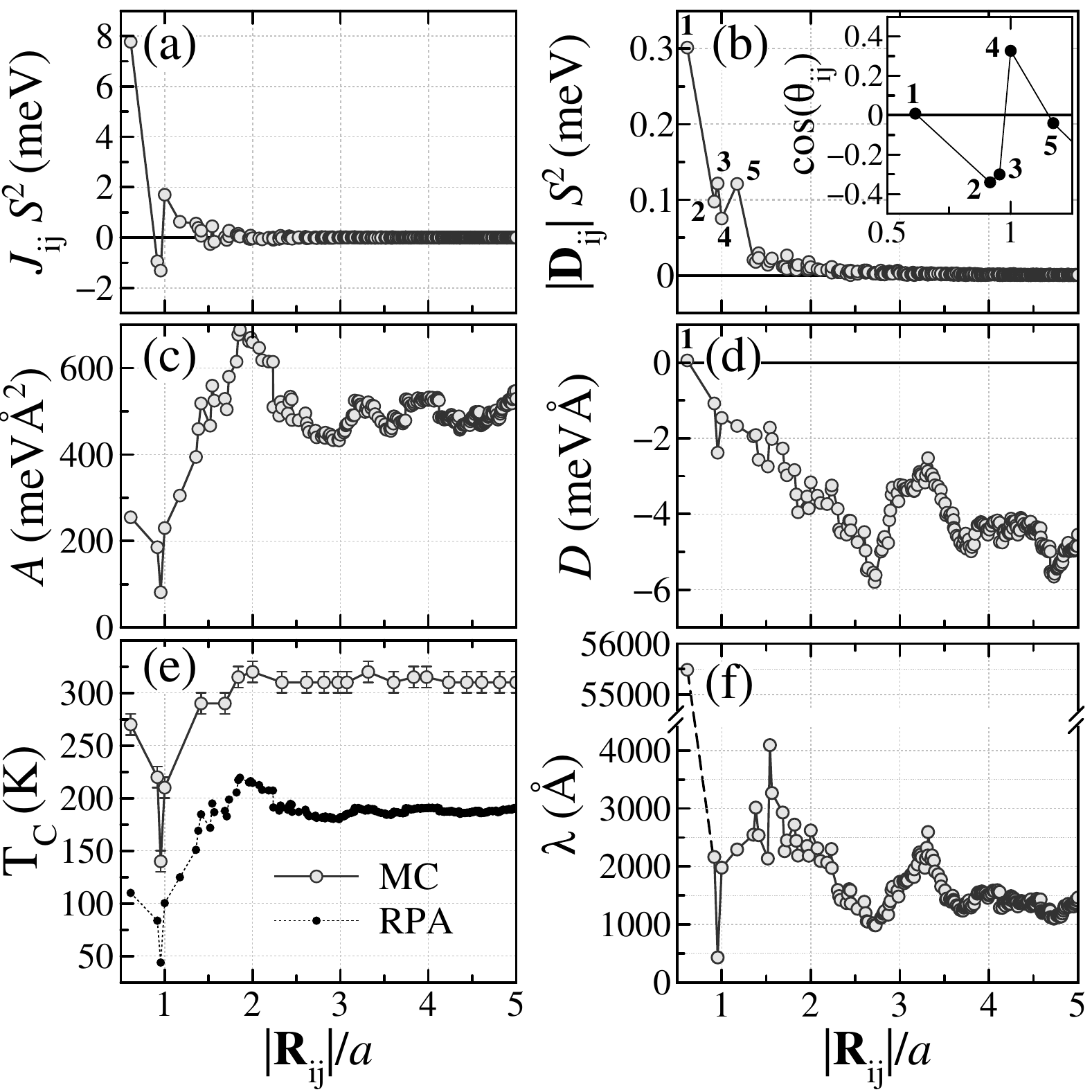}
 \caption{
(a) Exchange interaction parameters $J_{ij}$ and 
(b) absolute values of DM interaction vectors $\vert{\vn D}_{ij}\vert$ between Fe atoms, 
(c) micromagnetic spin stiffness $A$ and 
(d) spiralization $D$ (obtained via summation over all contributions up to $\vert{\vn R}_{ij}\vert$), 
(e) Curie temperature, $T_\text{C}$, and 
(f) period of the spin spirals, $\lambda$, as functions of the interatomic distance (between Fe atoms) $\vert{\vn R}_{ij}\vert$ (in units of the lattice parameter $a$).
Note that the parameters $J_{ij}$ and $\vert{\vn D}_{ij}\vert$ in (a) and (b) are multiplied by $S^2=\vert{\vn S}_i\vert\vert{\vn S}_j\vert$. The Curie temperature,
$T_\text{C}$, shown in (e) is computed from $J_{ij}$ parameters using Monte Carlo simulations (MC) and the random phase approximation (RPA). 
The inset in (b) presents $\cos(\theta_{ij})$ for the first five shells, where $\theta_{ij}$ is the angle between ${\vn R}_{ij}$ and ${\vn D}_{ij}$.
}
 \label{ss}
\end{figure}

Figure~\ref{ss}(a) depicts the exchange constants $J_{ij}$ as a function of the distance $\vert {\vn R}_{ij}\vert$ between two interacting magnetic moments. We included interaction pairs that are separated up to 5 lattice parameters, \textit{i.e.}, $R_\text{max}=5a$.
We observe that the $J_{ij}$'s decay rapidly with distance, and take the largest positive values for the first-nearest Fe-Fe neighbors. 
The interactions between second- and third-nearest neighbors exhibit $J_{ij}$'s of opposite, smaller value. 
Using this microscopic information, we evaluate the micromagnetic spin stiffness $A$ based on Eq.~\eqref{At}, the result of which is presented in Fig.~\ref{ss}(c) for an increasing number of considered interacting Fe pairs. 
As the individual contributions to $A$ follow the form $J_{ij}R_{ij}^2$, distant interaction partners constitute an important part of the micromagnetic spin stiffness. 
The decay of the $J_{ij}$ for larger distances competes with the quadratic increase of the separation $\vert {\vn R}_{ij}\vert$ between the moments, resulting in a diminishing oscillatory behavior of $A$ with respect to $\vert {\vn R}_{ij}\vert$.

To validate the computed $J_{ij}$ parameters, we use them to calculate the Curie temperature $T_\text{C}$. It is well known that the mean-field theory overestimates the Curie temperature, see for instance Refs.~\onlinecite{marjana2005, marjana2013}.
Therefore, we evaluate $T_\text{C}$ of the classical Heisenberg model considered in this work using either the random phase approximation (RPA) within a multi-sublattice approach~\cite{RPA, RPA1} or Monte Carlo simulations~\cite{monte}, which are both rather accurate but numerically more expensive methods.
In the Monte Carlo simulations we determine the Curie temperature from the peak of the temperature-dependent static susceptibility. 
In both methods we evaluate the Curie temperature of the feromagnetic state from different sets of calculations with an increasing number of pairs of magnetic moments that mutually interact.

The results summarized in Fig.~\ref{ss}(e) indicate that convergence of $T_\text{C}$
is achieved once we include interacting atoms that are further than 2.5 times the lattice parameter apart from each other.
Within RPA the computed $T_\text{C}$ for FeGe amounts to $187$~K, which is 33\% lower than the experimental value of $(280 \pm 2)$~K~\cite{Spencer2018}. 
In addition, we verified that accounting for the effect of DM interaction within RPA hardly affects this value of $T_\text{C}$. 
Using Monte Carlo simulations, we obtain a more accurate value for the Curie temperature $T_\text{C}$ of $(310\pm 10)$~K, which is only 12\% higher than the experimental value. Considering the electronic and magnetic complexity of the B20 FeGe, this is a very reasonable agreement, from which we conclude that the magnitude of the Heisenberg exchange constant and ultimately the spin stiffness are about 12\% too high.

Now, we turn to the detailed microscopic analysis of the tendency towards chiral magnetism in FeGe as mediated by the DM interaction.
Figure~\ref{ss}(b) displays the absolute value of the atomistic DM vectors
${\vn D}_{ij}$ as function of the interaction radius $R_{ij}$ between two Fe atoms.
Most prominently, we note that the magnitude of $\vn D_{ij}$ decays rapidly with distance, just like in the case of the $J_{ij}$, with the largest contribution originating from nearest-neighbor interactions.
Considering the factors ${\vn R}_{ij}\cdot{\vn D}_{ij}$ as integral contributions to the micromagnetic spiralization according to Eq.~\eqref{Dt}, we arrive at the slowly converging behavior of $D$ shown in Fig.~\ref{ss}(d). 
To understand this property, we introduce the angle $\theta$ between the orientation of $\vn D_{ij}$ and the associated bond connecting the interacting moments. Consequently, although the nearest neighbors provide large microscopic contributions to $\vn D_{ij}$, the associated angle $\theta$ amounts to nearly $90^\circ$, see inset of Fig.~\ref{ss}(b), rendering the overall effect on the micromagnetic spiralization negligible.
As the direction between bond vectors and microscopic DM vectors changes rapidly with distance, we find an oscillatory but slowly converging behavior of the micromagnetic DM parameter. The oscillatory behaviour of $A$ and $D$ with respect to $\vert{\vn R}_{ij}\vert$ has a strong effect on the period of the spin spiral $\lambda$, see Fig.~\ref{ss}(f).

\subsection{The Fermi-level dependence of the micromagnetic parameters}

\begin{figure}[t!]\center
\includegraphics[width=0.45\textwidth]{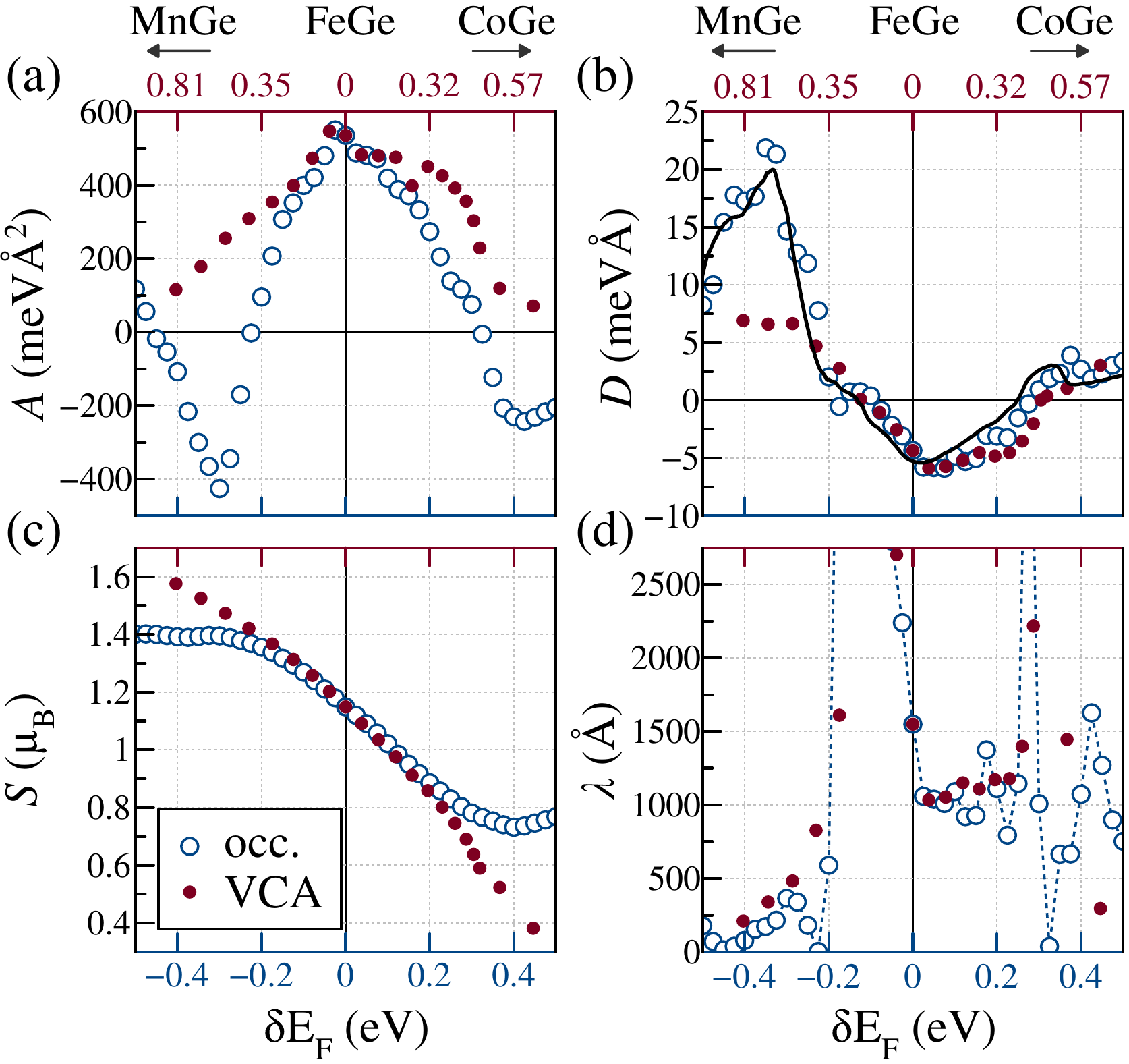}
 \caption{(Color online) 
 (a) Spin stiffness, $A$, (b) spiralization, $D$, 
 (c) magnetic moment of Fe atoms, $S$, and (d) period of the spin spirals, $\lambda$,
 as function of the band occupation (occ.) mediated by the Fermi energy shift $\delta E_\text{F}$ (open markers and bottom axis), and
 as function of the chemical composition $x$ 
 to Fe$_{1-x}$Mn$_{x}$Ge and Fe$_{1-x}$Co$_{x}$Ge compounds
 computed in the VCA (filled markers and top axis).
 Spin stiffness and spiralization are computed in accordance to equations~\eqref{At} and \eqref{Dt}, respectively. 
 $J_{ij}$ and ${\vn D}_{ij}$ were obtained by using the KKR method
 within LDA. The solid black line in figure (b) is the spiralization computed by utilizing a Berry phase approach (\fleur{}, LDA). 
 }
 \label{ef}
\end{figure}

Any real sample is subject to imperfections including impurities, anti-site defects, and off-stoichiometry, all of which can affect the filling of the electronic bands as well as the Fermi-surface topology. 
Likewise, correlation effects beyond those treated within LDA or GGA can modify the Fermi surface.
As the magnetic properties reflect immediately the spin and orbital nature of electrons near the Fermi level, variations of the latter can play a crucial role in correlating experiment and theory. 
Therefore, in this section we study how susceptible the magnetic properties of FeGe are with respect to variations of the band filling. Since the position of the Fermi level in the B20 magnet FeGe affects mainly band filling of the 3$d$-electrons, it can be associated with the replacement of Fe by Mn or Co atoms. Therefore, in addition to a simple shift of the Fermi level, we consider alloyed systems within the virtual crystal approximation (VCA) for a better comparison to experiment.
Within the method of infinitesimal rotations \cite{Liechtenstein:1987, Ebert:2009}, this merely amounts to shifting the upper integration limit of the convoluted Green's functions to values higher or lower than the Fermi level, obtaining band-filling-dependent interactions~\cite{Levaic, Mavropoulos}.

Fig.~\ref{ef} summarizes our results of the magnetic properties computed for different positions of the Fermi level (open markers) and chemical composition of FeGe alloys in VCA (filled markers).
Here, the micromagnetic parameters $A$ and $D$ were obtained as described in section~\ref{SS:Inf-rot_approach} by means of the KKR method, using the GGA lattice parameters and LDA characterising the magnetic interactions.
As can be seen from Fig.~\ref{ef}(a--d), for small changes of the Fermi level, the change of the parameters $A$, $D$, $S$, and $\lambda$ follows directly the change of the $d$-band filling of the alloys in the VCA, in agreement with the validity of the rigid-band approximation in this regime.
We find that a small change of the Fermi level by $-0.05$~eV ($+0.05$~eV) modifies the spin stiffness $A$ by $+2.7\%$ ($-8.9\%$) and the spiralization $D$ by $-28.9\%$ ($+33.3\%$). As a result, the period of the spin spiral changes drastically by $+44.5\%$ 
($-31.7\%$). The spin moment in Fig.~\ref{ef}(c) exhibits a linear change of the magnetic moments with band filling as expected from the Slater-Pauling curve.

We evaluate the Fermi-level dependence of the micromagnetic DM parameter also based on the Berry phase approach (see solid line in Fig.~\ref{ef}(b)), which agrees well with our KKR results. The obtained variation of the spiralization, mimicking the effect of doping, follows excellently recent theoretical and experimental reports for Fe$_{1-x}$Mn$_{x}$Ge ~\cite{Shibata:2013, Grigoriev:2013, Gayles:2015, Altynbaev2016} and Fe$_{1-x}$Co$_{x}$Ge~\cite{Spencer2018} alloys, 
for which the DM interaction changes sign at $x\approx0.2$ 
and $x\approx0.4$, respectively, leading to a divergence of the spin-spiral pitch, as shown in Fig.~\ref{ef}(d).
While the techniques and computer codes used in this work are substantially different, we find a good agreement between the results.

\section{Conclusions}\label{conclusions}

In conclusion, we carried out a comprehensive state-of-the-art DFT study of the magnetic properties of the prototypical B20 chiral magnet FeGe. 
Using different electronic-structure methods we determined and investigated both atomistic and micromagnetic parameters describing the exchange and DM interactions. They provide a consistent picture: In the absence of an external magnetic field and the neglect of the magnetic anisotropy conceived to be tiny due to cubic symmetry of the lattice, the ground state is found to be a helical spin texture following the same handedness as the crystal structure 
 with a period of $\lambda=(1450\pm 100)$~\AA{} obtained from the ratio of exchange spin stiffness and spiralization. While the handedness of the calculated spin spiral is consistent with experiment~\cite{Shibata:2013, Grigoriev:2013, Gayles:2015, Altynbaev2016, Spencer2018}, the pitch is about twice as large as for all reported measurements. In retrospect, this finding is consistent with previous theoretical studies on the magnetic properties of cubic FeGe, although the discrepancy of a factor two between theory and experiment was not further addressed in earlier work.

While we consider the experimentally determined period of FeGe as a hard experimental fact, confirmed by different experimental groups having taken different samples and having measured over a wide temperature range, the large discrepancy to the DFT results comes to a surprise considering (i) the predictive power of DFT on the pitch of DMI stabilized spin-spiral states proven for previous systems, like for a Mn monolayer on W(110)~\cite{Bode:07} to name one, and (ii) that the structural parameters obtained by total energy minimization, the local magnetic moment of Fe as well as the Curie temperature $T_\text{C}=(310\pm10)$~K agree well with experiments irrespective of the computational approach applied.

To deeper understand the origin of this discrepancy we explored the response of the period with respect to computational parameters like
the sampling of the BZ and the broadening of the Fermi distribution, structural details like lattice parameter and atomic positions, and the approximation of the exact exchange correlation functional employing LDA, GGA and LDA+$U$. 
We demonstrated that increasing the lattice parameter by 1\%, by simultaneously reducing the distance between the nearest Fe and Ge atoms by the same percentage results in a $\sim$8\% reduction of the spin spiral length. By reducing the strength of the vector portion of the exchange correlation (by the factor $\alpha$) or by applying the Hubbard-U correction to Ge $p$-orbitals, we were able to show that computed magnetic moment of Fe atoms (of 1.16$~\mu_\text{B}$ in GGA) can be tuned towards better agreement with the experimental value (of $\sim 1$~$\mu_\text{B}$). However, such a treatment of the exchange correlation functional increases the length of the spin spiral (by $\sim 5$\% if $U=1.5$~eV and $J=0.5$~eV or if $\alpha = 0.95\%$). In addition, we found that a small change of the electronic band filling around the Fermi level has a relatively strong influence on period of the spin spiral, which for instance, becomes 32\% shorter (74\% longer) when 3\% of Fe are substituted by Co (Mn).
Although the computational, structural, and correlation parameters and methods have a definite influence on the spin stiffness and spiralization of the B20 magnet FeGe, their effects are not sufficient to restore the experimental period of magnetic modulations in this compound, finally concluding that none of these parameters provide a convincing source for the failure of the DFT calculations. Since Fe in FeGe has a large magnetic moment of about 1~$\mu_\text{B}$, longitudinal fluctuations neglected in the present work can be excluded as possible source of error.

We suspect that the failure of DFT as a predictive tool has a more fundamental reason. We conjecture that the well-known and well-accepted micromagnetic relation between the spin stiffness $A$, the spiralization $D$, and the period $\lambda$ of the helical structure, $\lambda=4\pi |A/D|$, is not valid for FeGe. We speculate that it is not valid for any of the chiral B20 magnets. We consider the presence of higher-order magnetic interactions, such as the biquadratic, four-spin--three-site, four-spin--four-site exchange~\cite{Markus}, or the recently proposed topological-chiral interactions~\cite{Grytsiuk2019} as potentially relevant contributions that might violate the simple relation between the period and the micromagnetic parameters. 
A second possible reason for the remaining discrepancy roots in the fact that the ground state of FeGe could be a superposition of several helical spin-density waves propagating in the same direction but having different phases and different directions of the rotation axes~\cite{Dmitrienko12, Chizhikov12}.

Finally we would like to encourage new experiments where all three quantities, the period $\lambda$, the spin stiffness $A$ and the spiralization $D$, are measured independently under the same experimental conditions in order to verify or falsify the commonly accepted micromagnetic relation, $\lambda=4\pi |A/D|$.

\begin{acknowledgments}
We thank N.~Kiselev for the stimulation of this work and many discussions in the course of it. S.B. and Y.M.\ acknowledge support by the Deutsche Forschungsgemeinschaft (DFG) through the Collaborative Research Center SFB 1238 (Project C1) and Priority Programm SPP 2137. M.H., Y.M., and S.B.\ acknowledge funding from the DARPA TEE program through grant MIPR (\#HR0011831554) from DOI. We also gratefully acknowledge the J\"ulich Supercomputing Centre and RWTH Aachen University for providing computational resources under projects jara0161 and jiff40. 
\end{acknowledgments}


\begin{thebibliography}{9}
\bibitem{note-units}
Throughout this paper the scalar spin stiffness and spiralization constants, $A$ and $D$, are given in atomic units, which can be converted to SI units as used in experiments using the expression $\|\mathcal{A}\| \big[\text{pJ/m}\big] = \frac{1}{V_\text{uc}[\text{\AA}^3]}\frac{10^{10}\text{\AA}}{\text{m}} \frac{1.602\cdot 10^{-10}\text{pJ}}{\text{meV}} A \big[\text{meV\AA}^2\big] =0.0157 \cdot A \big[\text{meV\AA}^2\big] $ and $\|\mathcal{D}\| \big[\text{mJ}/\text{m}^2\big] = 0.157 \cdot D \big[\text{meV\AA}\big]$ taking into account
the GGA lattice parameter of $4.67$~\AA{}.
\bibitem{Bfield}
 The saturation magnetic field ${\vn B}$ corresponds to the Zeeman energy 
 $E = - {\vn m}\cdot {\vn B}$, where ${\vn m}$ is the total magnetic moment in the unit cell.
\bibitem{comment1}
 As we vary $u_\text{Fe}$, the change in the length $R_\text{Fe-Fe}$ of the shortest Fe-Fe bond amounts to less than $10\%$ of the change in $R_\text{Fe-Ge}$ such that we restrict our discussion to the latter.
\bibitem{monte}
 The Monte Carlo simulations included 3375 unit cells with a total of 13500 atoms.
 Performing one sampling event in every second sweep of the lattice, we initially relaxed the system in the first 500 steps before performing 4500 sampling events.
\end{thebibliography}
\end{document}